\documentclass[
    floatfix,
    pra,
    twocolumn,
    aps,
    superscriptaddress,
    showpacs,
    longbibliography
]{revtex4-1}
\usepackage{hyperref}
\hypersetup{
	colorlinks=true,
	citecolor=blue,
	linkcolor=blue,
	urlcolor=blue}
\usepackage{graphicx}
\usepackage[normalem]{ulem}
\usepackage{amsmath}
\usepackage{amsfonts}
\usepackage{amssymb}
\usepackage{epsfig}
\usepackage{subfigure}
\usepackage{mathtools}
\usepackage{braket}
\usepackage[usenames,dvipsnames]{color}
\usepackage{setspace}
\usepackage{bm}
\usepackage{xfrac}
\usepackage{times}
\usepackage{ulem}
\usepackage{dsfont}
\usepackage{xcolor}
\newcommand{\avg}[1]{\left\langle #1 \right\rangle}
\begin{document}
\title{Thermodynamics of Strongly Interacting Light Fields from Their Local Intensity Statistics}
\author{Love Kumar Sharma} \email{love.21phz0018@iitrpr.ac.in}
\affiliation{Department of Physics, Indian Institute of Technology Ropar, Rupnagar 140001, Punjab, India}
\author{Vishwa Pal}\email{vishwa.pal@iitrpr.ac.in}
\affiliation{Department of Physics, Indian Institute of Technology Ropar, Rupnagar 140001, Punjab, India}

\begin{abstract}
Optical thermodynamics describes the equilibrium of multimode nonlinear photonic systems, but its conventional Rayleigh-Jeans formalism becomes inadequate when nonlinear interactions substantially modify the modal occupations. Here, we develop a framework, based on exact relations, that connects local intensity statistics to the thermodynamics of interacting light. Three exact identities, valid for arbitrary coupling networks and local nonlinearities, relate the equation of state, mode-resolved nonlinear self-energies, and local intensity correlations. A self-consistent local intensity closure yields universal thermodynamic relations for Kerr interactions, describes saturable and competing cubic-quintic responses, and predicts equilibrium properties from the conserved power and energy of the launched field. The framework is benchmarked against exact one-dimensional transfer-operator solutions and lattice simulations in one and two dimensions.
\end{abstract}
\maketitle
%%%%%%%%%%%%%%%%%%%%%%%%%%%%%%%%%%%%%%%%%%%%%%%%%%%%%%%%%%%%%%%%%%%%%%%%%%%%%%%%%%%%%%%
%\section{Introduction}
\textit{Introduction} --- Light propagating in a multimode nonlinear photonic system, such as a multimode or multicore fiber, a waveguide array, or a lattice of coupled resonators, redistributes its power among the linear modes. Under suitable conditions, this redistribution drives the system towards a statistical equilibrium described by optical thermodynamics \cite{wu2019thermodynamic,parto2019thermodynamic,makris2020statistical, picozzi2014optical, wright2022physics}. In its conventional form, the equilibrium modal occupations of a highly multimodal nonlinear system follow the Rayleigh--Jeans (RJ) distribution, $n_k=T/(\varepsilon_k-\mu)$, where $\varepsilon_{k}$ is the linear-mode energy, while $T$ and $\mu$ are the optical temperature and optical chemical potential, determined by the conserved power and energy. This framework has successfully described several fundamental nonlinear phenomena such as Kerr beam self-cleaning \cite{krupa2017spatial,mangini2022statistical}, RJ condensation \cite{baudin2020classical,pourbeyram2022direct}, negative-temperature states \cite{marques2023observation,baudin2023observation}, calorimetry and Joule-Thomson expansion \cite{ferraro2024calorimetry,kirsch2025observation}, and spatiotemporal thermalization \cite{zanaglia2026spatiotemporal}. Beyond providing a framework for understanding equilibrium properties, nonlinear thermalization is also increasingly being explored as a means of controlling light, with applications in nonlinear imaging \cite{moussa2021spatiotemporal}, multimode fiber lasers \cite{wright2017spatiotemporal}, and passive optical routing \cite{dinani2025universal}. Furthermore, the nonlinear Schr\"odinger equation underlying these systems is closely related to models of Bose-Einstein condensates in optical lattices \cite{trombettoni2001discrete} and to fluids of light \cite{carusotto2013quantum}, which places the equilibrium thermodynamics of multimode photonic systems within the broader context of classical-field statistical mechanics.

Despite this progress, conventional optical thermodynamics largely treats equilibrium as that of an ideal gas of linear modes. Although nonlinear interactions drive the redistribution of optical power, their contribution to the conserved energy and their influence on the equilibrium modal occupations are not incorporated into the standard RJ description. This limitation becomes increasingly important when the nonlinear contribution to the energy is appreciable, particularly at high optical peak powers, or in systems with small modal spacing. 

Recent studies have made some progress in extending optical thermodynamics beyond the ideal RJ description. Using a transfer-integral operator (TIO) approach for a one-dimensional Kerr chain, Kabat \textit{et al.}~\cite{kabat2026optical} demonstrated that modal distributions can retain an RJ-like form with a renormalized chemical potential, $\mu_{\mathrm{eff}}=\mu-C\chi a$, where $C\simeq2$, together with a modified equation of state. The TIO method is, however, specific to one-dimensional chains, and these values were obtained at moderate interaction strength. Lian \textit{et al.}~\cite{lian2026nonlinear} obtained a van der Waals--type equation of state through a mean-field approximation, which assumes Gaussian field statistics and is therefore restricted to weak interactions. Related studies of the discrete nonlinear Schr\"odinger (DNLS) equation have provided statistical descriptions of nonlinear lattice systems, addressing equilibrium distributions, energy localization, breather formation, and condensation through Gibbsian, microcanonical, and mean-field approaches \cite{rasmussen2000statistical,rumpf2004simple,johansson2004statistical,gradenigo2021localization,iubini2025effective,giusfredi2026mean}, but have focused on localization and condensation rather than on the modal occupations measured in optical experiments. A framework is therefore still missing that holds for arbitrary coupling networks, dimensions, and local nonlinearities, remains valid when interactions suppress intensity fluctuations, and predicts the interacting equilibrium directly from the power and energy of the launched field.

In this work, we develop such a framework by identifying local intensity statistics as a central quantity governing the equilibrium thermodynamics of interacting light. Starting from generalized equipartition, we derive three exact relations that apply to arbitrary Hermitian coupling networks and local nonlinear responses. These identities connect the equation of state to local intensity correlations, relate each modal occupation to its nonlinear self-energy, and establish an exact sum rule for the weighted average self-energy. We then introduce a self-consistent closure for the local intensity distribution. For Kerr nonlinearities, this closure yields universal thermodynamic relations governed by a single dimensionless parameter, linking the nonlinear energy, intensity fluctuations, and compressibility. It also provides a route to optical thermometry using local intensity measurements, and describes saturable and competing cubic-quintic nonlinearities. We benchmark the theory against exact TIO solutions in one dimension and numerical simulations in both one and two dimensions. Finally, by combining the closure with the conserved power and energy of the launched field, we demonstrate how the equilibrium energy partition, temperature and modal distribution can be predicted directly from the initial conditions. Our results connect established RJ and modified-RJ descriptions to a more general thermodynamics of interacting optical fields, in which local nonlinear fluctuations play an explicit role. The framework also identifies the regime in which Gibbsian thermalization holds, and its extension to non-Gibbsian thermalization \cite{rasmussen2000statistical, iubini2013discrete, mithun2018weakly} is a natural next step.

\textit{Exact relations from generalized equipartition} --- We consider $M$ coupled complex amplitudes $\psi_m$ whose conservative evolution along the propagation distance $t$ is governed by \cite{christodoulides1988discrete}

\begin{equation}
i\dot{\psi}_m=\sum_n h_{mn}\psi_n+F'(A_m)\psi_m,\qquad A_m=|\psi_m|^2,
\label{eq:1}
\end{equation}

where $h_{mn}$ is a Hermitian coupling matrix and $F(A)$ is the local nonlinear energy. The conserved Hamiltonian and optical power are

\begin{equation}
H=\sum_{mn}h_{mn}\psi_m^*\psi_n+\sum_mF(A_m),\qquad P=\sum_mA_m.
\label{eq:2}
\end{equation}

We denote the linear and nonlinear contributions to the Hamiltonian by $H_{\rm L}$ and $H_{\rm NL}$, respectively. For a Kerr response, $F(A)=\chi A^2/2$, with $\chi>0$ corresponding to a defocusing nonlinearity in our energy convention. The linear supermodes $c_k=\sum_mU_{km}\psi_m$ obtained by diagonalizing the coupling matrix $h_{mn}$, give $H_{\rm L}=\sum_k\varepsilon_k|c_k|^2$. At thermal equilibrium, we consider the field to be described by the Gibbs measure $\rho\propto\exp[-\beta(H-\mu P)]$, with $T=\beta^{-1}$ as the thermodynamic temperature and $\mu$ the chemical potential \cite{rasmussen2000statistical,gradenigo2021localization}. We define the energy and power densities as $h=\avg{H}/M$ and $a=\avg{P}/M$, with $h_{\rm L,NL}=\avg{H_{\rm L,NL}}/M$.

Our starting point is the generalized equipartition theorem, which follows from an integration by parts in the complex field amplitudes \cite{tolman1918general, zhang2026equilibrium}. For any amplitude $u$, it gives (Appendix~\hyperref[Appendix_A]{A})

\begin{equation}
\Big\langle u^*\frac{\partial K}{\partial u^*}\Big\rangle=T,\qquad K= H-\mu P .
\label{eq:3}
\end{equation}

Applying this identity to the site amplitudes, $u=\psi_m$ and summing over all sites gives

\begin{equation}
h_{\mathrm{L}}-\mu a+\frac1M\sum_m\avg{A_mF'(A_m)}=T .
\label{eq:4}
\end{equation}

This exact optical virial relation connects the global thermodynamic variables to the local intensity statistics, without assuming weak interactions or Gaussian fluctuations. In terms of the compressibility factor $f\equiv\beta(h-\mu a)$, it can be equivalently written as

\begin{equation}
f=1-\frac{\beta}{M}\sum_m\avg{A_mF'(A_m)-F(A_m)} .
\label{eq:5}
\end{equation}

Thus, the departure from the ideal-gas value $f=1$ is governed by a local average nonlinear moment. Its sign is determined by $AF'(A)-F(A)$; for a defocusing Kerr response, this quantity is positive and the compressibility factor is reduced. In particular, since $AF'=2F$ for the Kerr nonlinearity, the relation simplifies to $f=1-\beta h_\mathrm{NL}$.

The same equipartition identity can be applied in the supermode basis, $u=c_k$, to obtain an exact relation for the modal occupations $n_k=\avg{|c_k|^2}$:

\begin{equation}
n_k=\frac{T}{\varepsilon_k-\mu+\Sigma_k},\qquad \Sigma_k=\frac{\mathrm{Re}\avg{c_k^*[F'(A)\psi]_k}}{n_k},
\label{eq:6}
\end{equation}

where $[F'(A)\psi]_k$ denotes the projection of the nonlinear polarization onto the $k$th supermode and $\Sigma_k$ is the equilibrium nonlinear self-energy of mode $k$. Equation (\ref{eq:6}) shows that the nonlinear interactions dress each linear mode by its own interaction-induced shift $\Sigma_k$. Consequently, the modal distribution retains the conventional RJ form at the thermodynamic temperature only when this dressing is independent of the mode index.

Although the nonlinear self-energy may vary from mode to mode, its occupation-weighted average is determined exactly by the local statistics.  Using the unitarity of the supermode transformation, we obtain

\begin{equation}
\bar\Sigma\equiv\frac{\sum_kn_k\Sigma_k}{\sum_kn_k}=\frac{1}{Ma}\sum_m\avg{A_mF'(A_m)} .
\label{eq:7}
\end{equation}

For the Kerr response, this becomes

\begin{equation}
\bar{\Sigma}=\chi a g_2,
\qquad
g_2=\frac{1}{Ma^2}\sum_m\left\langle A_m^2\right\rangle,
\label{eq:8}
\end{equation}

where $g_2$ is the normalized intensity correlation. The sum rule of Eq.~(\ref{eq:7}) thus provides a direct link between the microscopic intensity fluctuations and the average nonlinear dressing of the entire modal spectrum, and is independent of the underlying coupling network.

\begin{figure*}[htbp]
\includegraphics[width=17 cm]{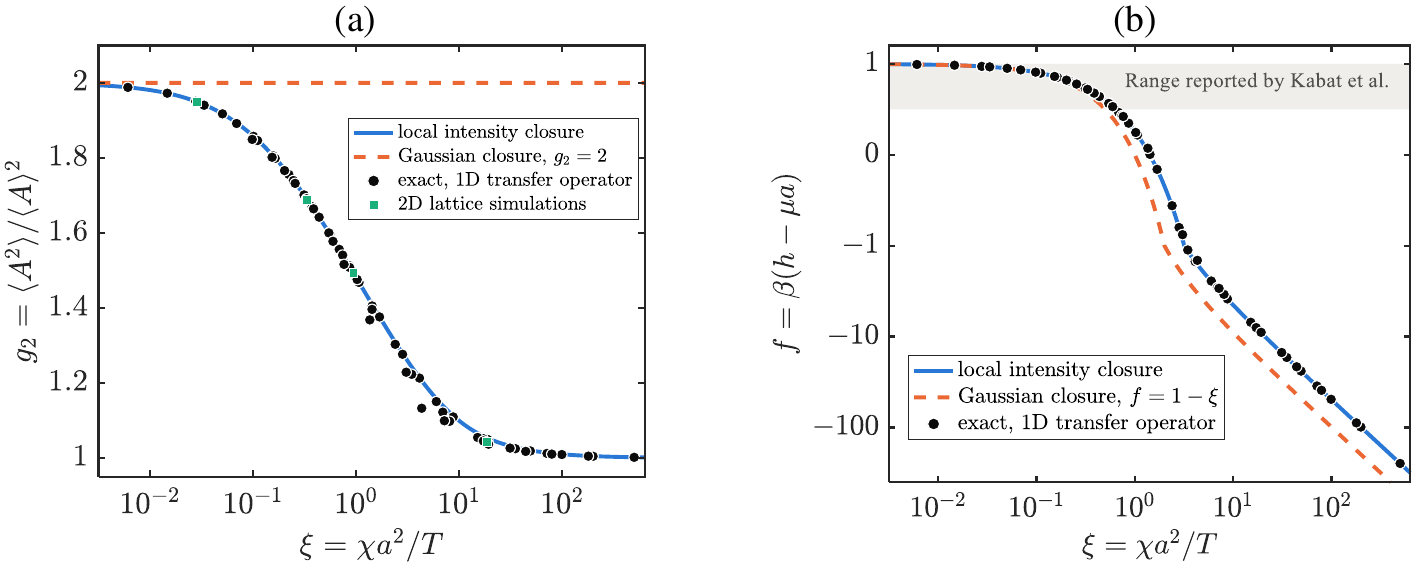} 
\caption{Universal thermodynamics of the Kerr lattice as a function of the dimensionless interaction parameter
$\xi=\chi a^2/T$. (a) Normalized local intensity correlation $g_2$ obtained from the local-intensity closure of Eq.~(\ref{eq:11}), compared with the Gaussian approximation ($g_2=2$), 63 exact equilibrium states of the one-dimensional (1D) chain, and simulations of the two-dimensional (2D) square lattice. (b) Compressibility factor $f$ obtained from the same closure, compared with the Gaussian approximation ($f=1-\xi$) and the exact 1D equilibrium states. The shaded band indicates the range reported in Ref.~\cite{kabat2026optical}.}
\label{fig_1}
\end{figure*}

When the self-energy is approximately mode-independent, $\Sigma_k\simeq\bar{\Sigma}$, the modal distribution reduces to a dressed RJ (dRJ) law with an effective chemical potential $\mu_{\mathrm{eff}}=\mu-\bar{\Sigma}$. For the Kerr nonlinearity, the chemical-potential shift is therefore

\begin{equation}
\mu-\mu_{\mathrm{eff}}=\chi a g_2=\frac{2h_{\mathrm{NL}}}{a}.
\label{eq:9}
\end{equation}

The shift in chemical potential is therefore governed by the local intensity correlation \(g_2\), rather than by a universal constant. For a Gaussian field, \(g_2=2\), whereas it approaches unity as repulsive nonlinear interactions suppress intensity fluctuations. This provides a microscopic interpretation of the coefficient \(C\simeq 2\) reported in Ref.~\cite{kabat2026optical}: \(C=g_2\). Likewise, Gaussian statistics in Eq.~(\ref{eq:4}) give the mean-field equation of state of Ref.~\cite{lian2026nonlinear}.

Equations~(\ref{eq:4}), (\ref{eq:6}), and (\ref{eq:7}) are exact and hold for arbitrary Hermitian coupling matrices and local nonlinearities, provided the Gibbs measure is normalizable.

\textit{Local closure and universal equation of state} --- The exact relations established above show that the nonlinear contribution to the equilibrium thermodynamics of a nonlinear multimode photonic system is governed by its local intensity statistics. For equivalent sites, the central quantity is the one-site intensity distribution $p(A)$. A Gaussian-field approximation fixes the normalized intensity correlation at $g_2=2$, independently of temperature. This description, however, cannot capture the suppression of intensity fluctuations at strong nonlinearity. At fixed mean power, the repulsive Kerr interaction increasingly suppresses the intensity variations as $T\rightarrow0$, driving $g_2$ towards its minimum value of unity. Consequently, the Gaussian approximation overestimates the nonlinear energy by up to a factor of two.

To describe this crossover, we introduce a self-consistent local-intensity closure,
\begin{equation}
p(A)=Z_A^{-1} \exp[-\beta F(A)+\beta\lambda A],
\qquad \langle A\rangle=a,
\label{eq:10}
\end{equation}

where $A\geq0$, $Z_A$ is the normalization constant, and the effective parameter $\lambda$ is determined by the mean-power constraint. For $\lambda=\mu$, this distribution is the exact Gibbs distribution of an isolated site \cite{rasmussen2000statistical,rumpf2004simple,iubini2025effective}. In mean-field descriptions, the local distribution is modified by an effective field representing the coupling to neighboring sites \cite{giusfredi2026mean}. Here, $\lambda$ plays a similar role, accounting approximately for the influence of the surrounding sites on the local intensity statistics. Unlike the mean-field theory of Ref.~\cite{giusfredi2026mean}, which decouples the hopping term and predicts $a$ and $h$ for given $(\beta, \mu)$, and unlike the isolated-site distribution with $\lambda=\mu$ used in Ref.~\cite{kabat2026optical}, this closure fixes $\lambda$ by the power alone and enters only through the local moments in the exact relations, Eqs.~(\ref{eq:4})--(\ref{eq:7}). Importantly, the contribution of the linear coupling to the global thermodynamics is retained through the exact relations derived above. The local closure reproduces the exact intensity distribution in the limits of infinite temperature, vanishing coupling, and vanishing nonlinearity.

For the Kerr response, $F(A)=\chi A^2/2$, the local distribution is a Gaussian truncated at $A=0$. Its thermodynamics is therefore governed by a single dimensionless parameter, $\xi=\chi a^2/T$, which measures the relative strength of nonlinear interactions and thermal fluctuations. Introducing $\zeta=\lambda/\sqrt{\chi T}$, we obtain (Appendix \hyperref[Appendix_B]{B})

\begin{equation}
\sqrt{\xi}=
\zeta+\frac{\varphi(\zeta)}{\Phi(\zeta)},
\qquad
g_2=\frac{1+\zeta\sqrt{\xi}}{\xi},
\qquad
f=\frac{1-\zeta\sqrt{\xi}}{2},
\label{eq:11}
\end{equation}

where $\varphi$ and $\Phi$ are the standard normal probability density and cumulative distribution functions, respectively. The first equation uniquely determines $\zeta$ for a given $\xi$, so that both the intensity correlation and the compressibility factor become universal functions of the interaction parameter. Thus, within this closure, all equivalent-site Kerr lattices share the same local-intensity statistics and equation of state, irrespective of their dimensionality and coupling geometry.

\begin{figure}[h!]
\includegraphics[width=8 cm]{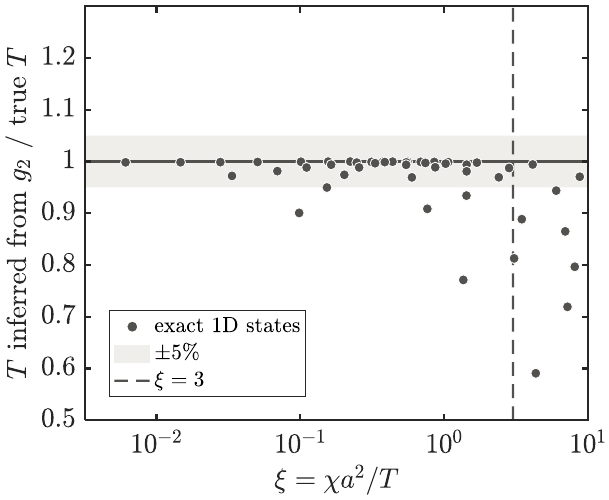}
\caption{Optical thermometry from local intensity statistics: Temperature inferred from the measured intensity correlation $g_2$ by inverting Eq.~(\ref{eq:11}), divided by the exact temperature, for the 48 exact 1D equilibrium states with $\xi\le10$, obtained with the TIO method. The shaded band marks $\pm5\%$; the dashed line marks $\xi=3$.}
\label{fig_2}
\end{figure}

In the weak-interaction regime, $\xi\ll1$, the Gaussian limit is recovered (Appendix \hyperref[Appendix_B]{B}), with $g_2=2-2\xi+O(\xi^2)$ and $f=1-\xi+O(\xi^2)$. Increasing $\xi$ drives the system away from Gaussian statistics, progressively suppressing local intensity fluctuations. At $\xi=2/\pi$, corresponding to $\zeta=0$, the closure gives $g_2=\pi/2$ and $f=1/2$, the value that Ref.~\cite{kabat2026optical} identified as the strong-nonlinearity limit of the compressibility factor. At stronger interactions, the thermodynamics enters a distinctly nonlinear regime in which $f$ becomes negative, while the intensity correlation approaches its lower limit. The corresponding strong-interaction asymptotics are

\begin{equation}
g_2=1+\frac{1}{\xi},
\qquad
f=\frac{1-\xi}{2},
\qquad \xi\gg1.
\label{eq:12}
\end{equation}

The range $1/2\leq f\leq 1$ and the bound $g_2\geq\pi/2$ of Ref.~\cite{kabat2026optical} therefore hold only for $\xi\leq 2/\pi$, and $C=g_2$ is close to $2$ only for $\xi\ll 1$.

\begin{figure*}[htbp]
\includegraphics[width=18 cm]{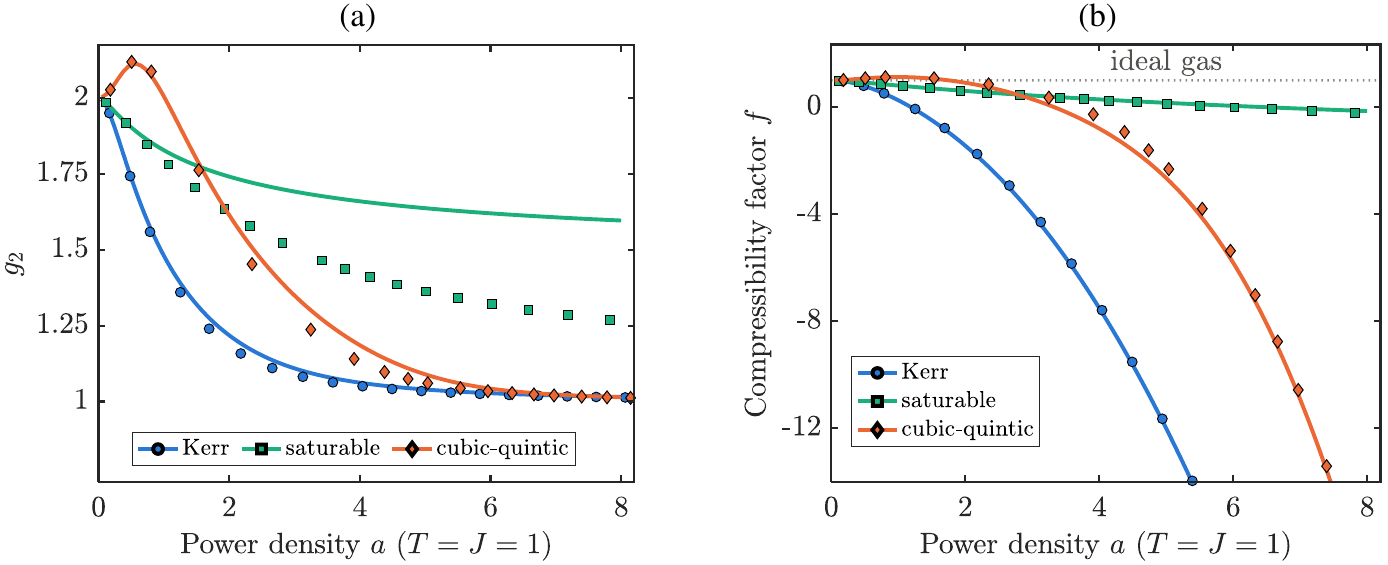}
\caption{Three local nonlinearities at fixed temperature, $T=J=1$: Kerr ($\chi=1$), saturable ($\chi=I_s=1$), and focusing cubic with defocusing quintic ($g_3=-0.5$, $g_5=0.1$). (a) Intensity correlation $g_2$  and (b) compressibility factor $f$ versus power density $a$. Symbols denote exact TIO solutions of the one-dimensional chain. Solid lines: local closure. The gray dotted line in (b) indicates the ideal-gas value $f=1$.}
\label{fig_3}
\end{figure*}

Figure~\ref{fig_1} benchmarks the universal relations against 63 exact equilibrium states of the 1D Kerr chain, obtained using the TIO method \cite{kabat2026optical,scalapino1972statistical,krumhansl1975dynamics} (Appendix~\hyperref[Appendix_C]{C}), and numerical simulations of the 2D square lattice. Without adjustable parameters, the exact values of $g_2$ collapse onto the predicted universal curve over five decades in $\xi$. The median deviation is $0.02\%$ for $\xi<1$, while the largest deviation for $\xi>10$ is $1.5\%$. The maximum deviation over the entire range is $6.5\%$, occurring in the crossover region around $\xi\simeq4$. The compressibility factor also follows the predicted curve, extending to negative values as low as $f\approx-250$, in the strong-nonlinearity limit. Remarkably, the 2D simulations fall on the same universal curves, supporting the applicability of the local closure beyond one dimension.

\textit{Optical thermometry from intensity histograms} --- The suppression of intensity fluctuations from $g_2=2$ towards $g_2=1$ is reminiscent of the crossover from a degenerate gas to a quasi-condensate in one-dimensional Bose gases \cite{kheruntsyan2003pair}, where density fluctuations have been used for thermometry \cite{esteve2006observations}. An analogous approach provides a means of measuring the thermodynamic state of multimode light. For a spatially homogeneous output, the intensity histogram obtained from a single site-resolved output image of a waveguide array \cite{lederer2008discrete,silberberg2009universal} or multicore fiber provides the local correlation $g_2$. Inverting Eq.~(\ref{eq:11}) then determines $\xi$ and, for known $\chi$ and mean power $a$, the temperature $T$, without requiring a modal decomposition. Applied to the exact 1D equilibrium states, this procedure predicts the temperature with a median error of $0.3\%$ for $\xi<3$ (Fig.~\ref{fig_2}). At larger $\xi$, the coupling between neighboring sites suppresses the fluctuations further than a one-site closure allows, and the inferred temperature is systematically too low. Local intensity statistics thus provide a practical observable for optical thermometry. The intensity correlation also carries information about the nature of the thermalized state. For a defocusing Kerr lattice, $g_2$ remains below its Gaussian value of $2$ at positive temperature, reaching it only at the infinite-temperature boundary, $h=\chi a^2$. Values above $2$ therefore signal non-Gibbsian behavior associated with breather formation \cite{rasmussen2000statistical, iubini2013discrete}.

\textit{Beyond Kerr: saturable and cubic-quintic nonlinearities} --- The exact identities and local-intensity closure developed above are not restricted to Kerr interactions. For a general local response, the required moments are evaluated through one-dimensional quadratures using the corresponding nonlinear energy $F(A)$. If $F(A)$ is homogeneous of degree $p$ in intensity, such that $AF'(A)=pF(A)$, the virial relation and equation of state reduce to $h_{\mathrm L}+p h_{\mathrm{NL}}-\mu a=T$ and
$f=1-(p-1)\beta h_{\mathrm{NL}}$. Two responses of practical importance are not homogeneous. A saturable response, $F'(A)=\chi A/(1+A/I_s)$, describes photorefractive lattices \cite{fleischer2003observation}, and for it

\begin{equation}
AF'-F=\chi I_s^2\left[
\ln\left(1+\frac{A}{I_s}\right)
-\frac{A/I_s}{1+A/I_s}
\right].
\label{eq:13}
\end{equation}

It grows only logarithmically at large intensity. Moreover, $F'(A)$ approaches the constant $\chi I_s$, so that the
nonlinear response increasingly acts as an approximately uniform shift of the mode frequencies. Consequently, the departure from the ideal equation of state remains comparatively small.

\begin{figure*}[htbp]
\includegraphics[width=18 cm]{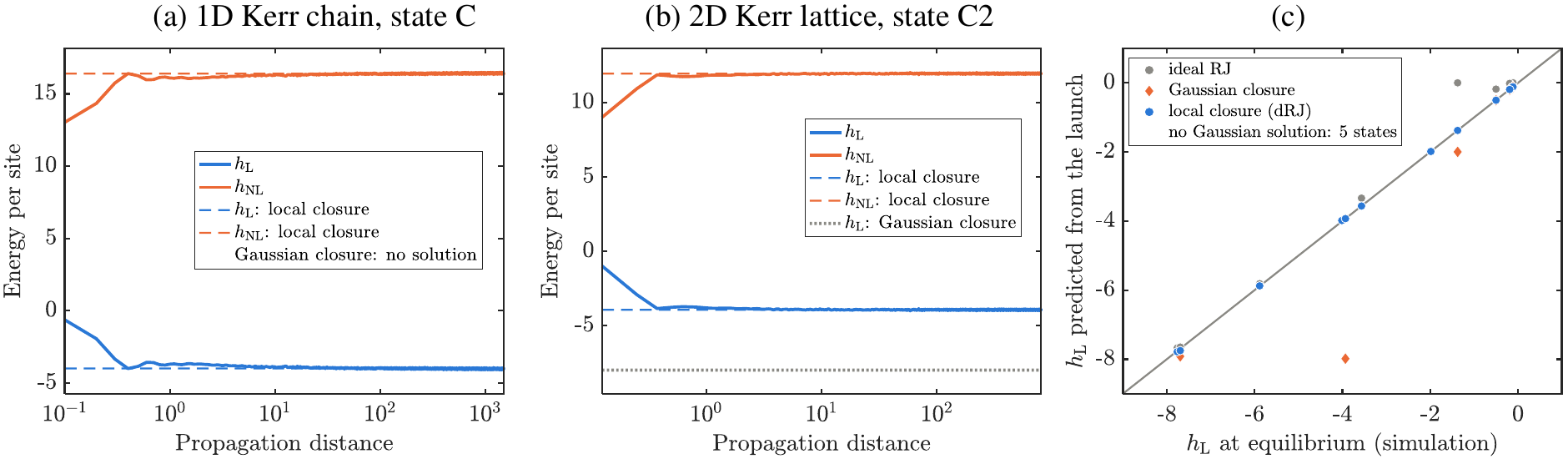}
\caption{Prediction of equilibrium from conserved launch conditions. (a), (b) Thermalization of the 1D state C and the 2D state C2: linear and nonlinear energy densities during the propagation (solid) and local-closure predictions from the launch values $(h,a)$ (dashed). Dotted line in (b): Gaussian closure prediction of $h_{\rm{L}}$; for state C the Gaussian closure has no solution. The horizontal axis is logarithmic. (c) Predicted versus simulated equilibrium linear energy for twelve one-and two-dimensional Kerr and one-dimensional saturable states. The Gaussian closure has no positive-temperature solution for five of them. The ideal theory keeps $h_{\mathrm{L}}$ at its launch value, and therefore misses the redistribution of energy between $h_{\rm{L}}$ and $h_{\rm{NL}}$.}
\label{fig_4}
\end{figure*}

Another nonlinear response we considered is a cubic-quintic response, $F'(A)=g_3A+g_5A^2$ with $g_3<0<g_5$, which describes an attractive interaction at low intensity and a repulsive one at high intensity, similar to the setting of liquid-light condensates \cite{michinel2002liquid}. The positive quintic term ensures normalizability of the Gibbs measure for positive temperature.

Figure~\ref{fig_3} compares the local closure with exact TIO solutions of the 1D chain for the three nonlinearities. For the cubic-quintic chain, the attraction raises $f$ above 1 and makes the intensity statistics super-Gaussian, $g_2>2$, at low power, while at high power $f$ becomes large and negative. The closure follows the exact $g_2$ within 5\% and the exact $f$ within 0.5 throughout, whereas the Gaussian closure (not shown) overestimates the departure from ideality, $1-f$, by up to a factor of 13. For the saturable chain the departure from ideality stays small, and the closure gives $f$ within $0.2$. The %one-site
local-intensity closure, however, overestimates $g_2$ at high power, by about $25\%$ at the highest powers in Fig. \ref{fig_3}, because a saturable on-site potential is soft and the amplitude fluctuations are set mainly by the coupling between sites.
%: a saturable on-site potential is soft, so that the amplitude fluctuations are limited mainly by the coupling between sites, which a one-site description does not contain.
The exact relations themselves hold in all cases; Table \ref{tab_1} (Appendix \hyperref[Appendix_E]{E}) tests Eq.~(\ref{eq:4}) on the exact solutions.

\textit{Predicting equilibrium from the launched field} --- The local-intensity closure also provides a route for predicting the equilibrium state directly from the conserved power and energy of the launched field. Unlike conventional approaches, in which the temperature and chemical potential are determined by fitting the final modal distribution, our approach obtains the equilibrium temperature and effective chemical potential from the initial conditions. For a given linear spectrum and conserved power and energy densities, \(a\) and \(h\), the equilibrium temperature is obtained from the energy constraint,
\begin{equation}
h=h_{\mathrm{L}}^{\rm dRJ}(T,a)+h_{\mathrm{NL}}^{\mathrm {cl}}(T,a),
\label{eq:14}
\end{equation}

where $h_{\mathrm{L}}^{\mathrm{dRJ}}$ is the linear energy density evaluated using the dressed RJ distribution $n_{k}={T}/{({\varepsilon_{k}}-\mu_{\mathrm{eff}})}$, i.e., Eq.~(\ref{eq:6}) with $\Sigma_k\simeq{\bar\Sigma}$ and  $\mu_{\rm{eff}}$ fixed by the power $a$. Similarly, $h_{\mathrm{NL}}^{\rm cl}$ is the nonlinear energy density obtained from the local-intensity distribution of Eq.~(\ref{eq:10}), for the Kerr response, $h_{\rm NL}^{\rm cl}=\chi a^2g_2(\xi)/2$. The explicit constraints and the solution procedure are given in Appendix~\hyperref[Appendix_D]{D}.

\begin{figure*}[htbp]
\includegraphics[width=18cm]{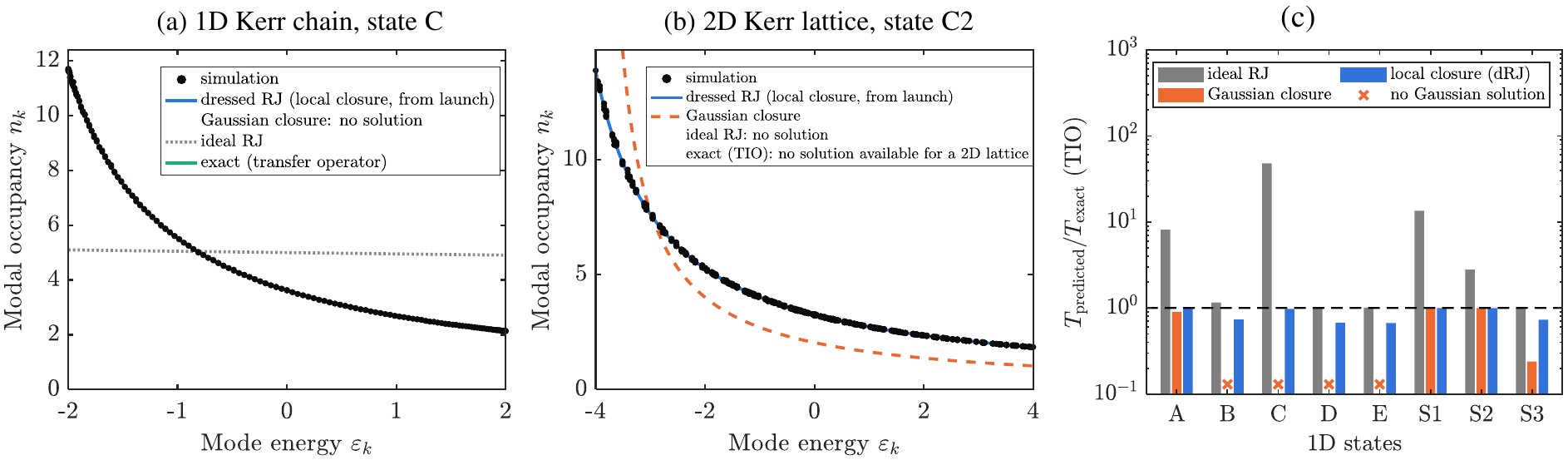}
\caption{Prediction of modal occupations and temperature from the conserved launch conditions. (a) One-dimensional Kerr chain, state C, with 256 modes. (b) Two-dimensional Kerr lattice, state C2, with 1024 modes. Symbols show the simulated modal occupations; solid curves show the dressed RJ (dRJ) prediction obtained from the conserved launch energy and power. In (a), the exact TIO occupations for the same $(h,a)$ coincide with the dRJ curve, and the ideal RJ theory (dotted), which keeps $h_{\mathrm{L}}$ at its launch value, gives an almost uniform distribution; the Gaussian closure has no solution for state C. In (b), the Gaussian closure is shown dashed. (c) Temperature predicted from the launch, divided by the exact TIO temperature for the same $(h,a)$, for the 1D Kerr (A--E) and saturable (S1--S3) states. The dashed line marks exact agreement; crosses mark states for which the Gaussian closure has no positive-temperature solution.}
\label{Fig_5}
\end{figure*}

We test this prediction using numerical propagation of one-dimensional chains of 256 sites and two-dimensional square lattices of $32\times32$ sites, considering Kerr and saturable nonlinearities. The initial fields have uniform intensity and controlled phase disorder, and hence $g_2=1$ at launch. %(Appendix~\hyperref[Appendix_E]{E}).
During propagation, energy is redistributed between the linear and nonlinear contributions while the total Hamiltonian and optical power remain conserved. Figures \ref{fig_4}(a) and \ref{fig_4}(b) show this redistribution for states C and C2: the linear energy falls by about four units per site, from $-0.10$ to $-4.01$ in C and from $0.02$ to $-3.93$ in C2, and the nonlinear energy rises by the same amount. The local closure predicts both plateaus from the launch values, whereas the Gaussian closure predicts for C2 twice the change in the linear energy and has no solution for C. Across twelve launch conditions, the local closure predicts the equilibrium linear energy to within $1\%$ of the numerical results [Fig. \ref{fig_4}(c), Table \ref{tab_2}]. These results demonstrate that the local intensity statistics provide the essential correction required to predict the nonlinear energy partition from the initial conditions.

The importance of this correction becomes evident when compared with conventional approximations. The ideal RJ description neglects the nonlinear contribution to the conserved energy, whereas a Gaussian closure fixes the intensity correlation at $g_2=2$, irrespective of the interaction strength. Consequently, the ideal RJ description keeps $h_\mathrm{L}$ at its launch value and misses the energy redistribution of up to four units per site in Figs. \ref{fig_4}(a) and \ref{fig_4}(b). The Gaussian closure fails for five states. For state C, as an example, the Gaussian closure requires $h_{\rm{L}}=h-\chi a^2=-12.6$, below the lower bound $a\varepsilon_{\rm{min}}=-10$ of any field with power $a$.

\textit{Modal occupations and temperature} --- Beyond the energy partition, the same construction predicts the equilibrium occupations of individual linear modes. We demonstrate this for the one-dimensional Kerr state C and the two-dimensional state C2 [Figs.~\ref{Fig_5}(a) and \ref{Fig_5}(b)]. The predicted dressed RJ distributions reproduce the simulated occupations of all 256 and 1024 modes, respectively, with median deviations of $0.7\%$ and $1.3\%$. For state C, the prediction also agrees with the exact TIO occupations at the same $(h, a)$, to within $0.9\%$ for every mode. Thus, the local intensity statistics determine both the nonlinear correction to the energy and the dressing of the modal spectrum, allowing the equilibrium distribution to be predicted without fitting the final modal occupations. For the 1D states A, C, S1, and S2, the predicted temperature lies within $2.5\%$ of the exact TIO temperature at the same $(h, a)$ (C: $10.56$ against $10.83$) [Fig.~\ref{Fig_5}(c)]. In the nearly condensed states B, D, E, and S3, for which $\xi>3$, it lies $27\%$--$33\%$ below it, as does the temperature of an RJ fit to the simulated occupations, because the occupations there are no longer of the dressed RJ form; the energy partition is still predicted within $1\%$. The ideal RJ theory comes close to the exact temperature of these states [Fig. \ref{Fig_5}(c)] only because their linear energy changes little during propagation.

%Predicting the temperature of these strongly interacting states requires going beyond the one-site closure; this work is in progress and will be presented in a future publication.

\textit{Experimental implications} --- The predictive construction also has a direct experimental interpretation. The launched optical power determines $a$, while the initial intensity and modal distribution determine the conserved energy density $h$. These quantities, together with the linear spectrum and nonlinear response, are sufficient to determine the equilibrium thermodynamic parameters. For Kerr interactions, the parameter $\xi=\chi a^2/T$ characterizes the relative strength of nonlinear interactions and thermal fluctuations. It is the ratio of the nonlinear frequency shift $\chi a$ to the thermal energy per unit power, $T/a$: small for incoherent, speckle-like inputs and large for inputs concentrated in a low-order supermode, where beam self-cleaning and condensation are relevant. Consequently, controlling the input modal composition provides a means of preparing and predicting different equilibrium states of multimode optical systems. The closure assumes equivalent sites, as in waveguide arrays and multicore fibers; for graded-index multimode fibers, whose transverse potential makes the local power position-dependent, $\lambda$ in Eq.~(\ref{eq:10}) would have to vary with position, whereas the exact relations, Eqs.~(\ref{eq:4})--(\ref{eq:7}), apply unchanged to any discretization of the transverse field.

\textit{Conclusion and outlook} --- We have established a statistical-mechanical framework for interacting multimode optical fields beyond the ideal Rayleigh--Jeans approximation. Three exact identities connect local intensity statistics, the equation of state, and mode-resolved nonlinear self-energies for arbitrary coupling networks and local nonlinearities. Building on these identities, a self-consistent local-intensity closure yields a universal equation of state for Kerr systems, governed by the single dimensionless parameter $\xi=\chi a^2/T$, revealing the role of intensity fluctuations in determining modal dressing and compressibility, and it describes saturable and competing cubic-quintic responses with the same quadrature. Importantly, the framework predicts the equilibrium energy partition and, when the nonlinear self-energy is approximately mode-independent, the modal occupations directly from the conserved launch power and energy. Agreement with exact TIO solutions and one- and two-dimensional lattice simulations demonstrates its predictive accuracy. In addition, the local intensity correlation provides a direct route to optical thermometry, potentially allowing the equilibrium temperature to be inferred from a site-resolved intensity measurement without reconstructing the modal distribution. These results establish local intensity statistics as a fundamental link between nonlinear interactions and the equilibrium thermodynamics of classical optical fields. Extending this framework to non-Gibbsian thermalization, where the Gibbs measure no longer applies, is a natural next step.

\textit {Acknowledgments} --- We acknowledge the funding assistance from the Science and Engineering Research Board, India (Grant No. CRG/2021/003060) and Indian Institute of Technology Ropar (Grant No.  9-230/2018/IITRPR/3255). Love Kumar Sharma thanks IIT Ropar for the financial support.

\appendix
\setcounter{equation}{0}
\renewcommand{\theequation}{A\arabic{equation}}
\section*{A\lowercase{ppendix} A: D\lowercase{erivation of the exact relations}}
\label{Appendix_A}
\textit{Generalized equipartition.}--- Let $u=x+iy$ denote any complex field amplitude, with $\partial_{u^*}=(\partial_x+i\partial_y)/2$. Treating $u$ and $u^*$ as independent variables, integration by parts gives \cite{tolman1918general}

\begin{equation}
0=\int\mathcal D\psi\,
\frac{\partial}{\partial u^*}
\left[u^*e^{-\beta K}\right],
\label{eq:A1}
\end{equation}

where the boundary term vanishes for a normalizable Gibbs distribution. Expanding the derivative and dividing by the partition function yields

\begin{equation}
\left\langle u^*\frac{\partial K}{\partial u^*}\right\rangle
=\frac{1}{\beta}=T.
\label{eq:A2}
\end{equation}

\textit{Optical virial relation and equation of state.}---
For the site amplitudes, differentiation of $K=H-\mu P$ gives

\begin{equation}
\frac{\partial K}{\partial\psi_m^*}
=\sum_nh_{mn}\psi_n+
[F'(A_m)-\mu]\psi_m.
\label{eq:A3}
\end{equation}

Multiplying by $\psi_m^*$, taking the ensemble average, and summing over all sites gives

\begin{equation}
MT=\langle H_{\mathrm L}\rangle-\mu\langle P\rangle
+\sum_m\langle A_mF'(A_m)\rangle,
\label{eq:A4}
\end{equation}

which, after division by $M$, gives Eq.~(\ref{eq:4}). Using $h=h_{\mathrm L}+h_{\mathrm{NL}}$, we further obtain
\begin{equation}
h-\mu a=T-\frac{1}{M}\sum_m
\left\langle A_mF'(A_m)-F(A_m)\right\rangle,
\label{eq:A5}
\end{equation}
which gives the compressibility factor in Eq.~(\ref{eq:5}).

\textit{Modal occupations and nonlinear self-energy sum rule.}--- In the supermode representation, the derivative of the grand-canonical Hamiltonian is

\begin{equation}
\frac{\partial K}{\partial c_k^*}
=(\varepsilon_k-\mu)c_k+[F'(A)\psi]_k,
\end{equation}

where $[F'(A)\psi]_k =\sum_m U_{km}F'(A_m)\psi_m.$

\hspace{4pt}

Applying generalized equipartition gives

\begin{equation}
(\varepsilon_k-\mu)n_k+
\operatorname{Re}\left\langle
c_k^*[F'(A)\psi]_k\right\rangle=T,
\label{eq:A6}
\end{equation}

where $n_k=\langle|c_k|^2\rangle$. Dividing by $n_k$ and using the definition of $\Sigma_k$ yields the modal occupation identity in Eq.~(\ref{eq:6}).

The self-energy sum rule follows from the completeness of the unitary transformation:

\begin{align}
\sum_k n_k\Sigma_k
&=\operatorname{Re}\sum_k
\left\langle c_k^*[F'(A)\psi]_k\right\rangle
\nonumber\\
&=\operatorname{Re}\sum_m
\left\langle\psi_m^*F'(A_m)\psi_m\right\rangle
\nonumber\\
&=\sum_m\left\langle A_mF'(A_m)\right\rangle,
\label{eq:A8}
\end{align}

which establishes Eq.~(\ref{eq:7}). Since $\sum_k n_k=\langle P\rangle=Ma$, division by the total occupation gives the occupation-weighted average self-energy.

\textit{Kerr specialization.}---
For $F(A)=\chi A^2/2$, one has $F'(A)=\chi A$, and Eq. (\ref{eq:7}) gives Eqs. (\ref{eq:8}) and (\ref{eq:9}) with $h_{\mathrm{NL}}={\chi}\sum_m\langle A_m^2\rangle/2M$. This relation determines the occupation-weighted average self-energy exactly; it does not imply that $\Sigma_k$ is independent of $k$. Replacing the individual self-energies by ${\bar\Sigma}$ in the exact relation for the modal occupancies (Eq.~\ref{eq:6}) requires the additional approximation of a sufficiently mode-independent self-energy. Only in this limit does the dressed Rayleigh--Jeans description follow.

\setcounter{equation}{0}
\renewcommand{\theequation}{B\arabic{equation}}
\section*{A\lowercase{ppendix} B: L\lowercase{ocal closure and \uppercase{K}err moments}}
\label{Appendix_B}

\textit{Truncated-Gaussian distribution.}---
For the Kerr nonlinearity, $F(A)=\chi A^2/2$, the local distribution in Eq.~(\ref{eq:10}) becomes

\begin{equation}
p(A)=Z_A^{-1}
\exp\left(-\frac{\beta\chi A^2}{2}+\beta\lambda A\right),
\qquad A\geq0.
\label{eq:B1}
\end{equation}

Introducing $x=\sqrt{\beta\chi}\,A$ and $\zeta=\lambda\sqrt{\beta/\chi}$, its normalization is

\begin{align}
Z_x
&=\int_0^\infty e^{-x^2/2+\zeta x}\,\mathrm dx\nonumber\\
&=\sqrt{2\pi}\,e^{\zeta^2/2}\Phi(\zeta),
\label{eq:B2}
\end{align}

where $\Phi$ denotes the standard normal cumulative distribution. Writing $\varphi$ for the corresponding probability density, differentiation with respect to $\zeta$ gives

\begin{equation}
\langle x\rangle
=\frac{\partial\ln Z_x}{\partial\zeta}
=\zeta+\frac{\varphi(\zeta)}{\Phi(\zeta)}.
\label{eq:B3}
\end{equation}

The mean-power constraint $\langle A\rangle=a$ therefore relates the scaled mean to $\xi=\chi a^2/T$ through $\langle x\rangle=\sqrt{\xi}$.

The second moment follows directly by integrating the derivative of $xe^{-x^2/2+\zeta x}$ over the positive real axis:

\begin{equation}
0=\int_0^\infty
\frac{\mathrm d}{\mathrm dx}
\left(xe^{-x^2/2+\zeta x}\right)\mathrm dx.
\label{eq:B4}
\end{equation}

After normalization, this gives

\begin{equation}
\langle x^2\rangle=1+\zeta\langle x\rangle
=1+\zeta\sqrt{\xi}.
\label{eq:B5}
\end{equation}

Using $A=x/\sqrt{\beta\chi}$, the second intensity moment is consequently obtained from

\begin{equation}
\frac{\langle A^2\rangle}{a^2}
=\frac{\langle x^2\rangle}{\xi}.
\label{eq:B6}
\end{equation}

Together with the Kerr nonlinear energy $h_{\mathrm{NL}}=\chi\langle A^2\rangle/2$, this gives the universal expressions, Eq.~(\ref{eq:11}), reported in the main text.

\textit{Limiting behavior.}---
In the weak-interaction limit, $\xi\ll1$, the intensity distribution approaches the exponential form. Expansion of the moment relations gives

\begin{equation}
g_2=2-2\xi+O(\xi^2),
\qquad
f=1-\xi+O(\xi^2).
\label{eq:B7}
\end{equation}

At $\zeta=0$, one obtains $\langle x\rangle=\sqrt{2/\pi}$, corresponding to $\xi=2/\pi$, $g_2=\pi/2$, and $f=1/2$.

In the opposite limit, $\xi\gg1$, the distribution is narrowly concentrated around its mean and the lower truncation becomes negligible. The leading asymptotic behavior is

\begin{equation}
g_2=1+\frac{1}{\xi}+O(\xi^{-2}),
\qquad
f=\frac{1-\xi}{2}+O(\xi^{-1}),
\label{eq:B8}
\end{equation}

with the intensity variance approaching $\mathrm{Var}(A)\simeq T/\chi$. Thus, the limiting moments follow directly from the truncated-Gaussian distribution and its normalization.

\setcounter{equation}{0}
\renewcommand{\theequation}{C\arabic{equation}}
\section*{A\lowercase{ppendix} C: E\lowercase{xact transfer-integral solution of the \uppercase{1D} chain}}
\label{Appendix_C}

\textit{Transfer-integral kernel.}---
For a one-dimensional periodic chain with nearest-neighbor coupling $J$, we write $ \psi_m=\sqrt{A_m}e^{i\theta_m}$ and follow the transfer-integral approach of Refs. \cite{scalapino1972statistical, krumhansl1975dynamics}, as applied to the DNLS lattice in Refs. \cite{kabat2026optical, rasmussen2000statistical}. The grand-canonical weight contains the local factor $ w(A)=\exp[-\beta F(A)+\beta\mu A]$ and the bond factor
$\exp\left[2\beta J\sqrt{AA'}\cos(\theta'-\theta)\right]$. Expanding the angular dependence in Fourier harmonics, $e^{x\cos\theta} =\sum_{\ell}I_\ell(x)e^{i\ell\theta}$, with $I_\ell$ the modified Bessel functions, gives the radial kernels

\begin{equation}
\mathcal K_\ell(A,A')
=\sqrt{w(A)}\,
I_\ell(2\beta J\sqrt{AA'})
\sqrt{w(A')}.
\label{eq:C1}
\end{equation}

The partition function of a periodic chain of $M$ sites is proportional to $\operatorname{Tr}(\mathcal K_0^M)$. With $\lambda_0$ and $y_0(A)$ the largest eigenvalue and the normalized eigenfunction of $\mathcal K_0$, the one-site distribution in the thermodynamic limit is $p(A)=y_0(A)^2$. The quantities $a$ and $h-\mu a$ follow from $\ln\lambda_0$ by differentiation, at fixed $\beta$ and $\mu$, respectively:
\begin{equation}
a=\frac{\partial\ln\lambda_0}{\partial(\beta\mu)},
\qquad
h-\mu a=-\frac{\partial\ln\lambda_0}{\partial\beta}.
\label{eq:C2}
\end{equation}

\textit{Linear energy from the bond statistics.}---
The bond marginal $P(A,A')= \frac{y_0(A)\mathcal K_0(A,A')y_0(A')}{\lambda_0}$ gives the linear energy independently of Eq.~(\ref{eq:4}), since the average of $\cos(\theta'-\theta)$ at fixed amplitudes is $I_1/I_0$:

\begin{equation}
h_{\mathrm L}
=-2J\left\langle
\sqrt{AA'}\,
\frac{I_1(2\beta J\sqrt{AA'})}
     {I_0(2\beta J\sqrt{AA'})}
\right\rangle_P .
\label{eq:C3}
\end{equation}

Table \ref{tab_1} compares this value with $T+\mu a-\langle AF'(A)\rangle$ from Eq.~(\ref{eq:4}).

\textit{Modal occupations.}---
The field correlation $C(r)=\langle\psi_0^*\psi_r\rangle$ is carried by the first angular harmonic \cite{scalapino1972statistical}. With $v(A)=\sqrt{A}\,y_0(A)$ and the eigenvalues $\lambda_{\nu,1}$ and eigenfunctions $y_{\nu,1}$ of $\mathcal K_1$,

\begin{equation}
C(r)=\sum_\nu
\left|\int_0^\infty v\,y_{\nu,1}\,\mathrm dA\right|^2
\left(\frac{\lambda_{\nu,1}}{\lambda_0}\right)^{|r|}.
\label{eq:C4}
\end{equation}

The exact occupations are then obtained from $n_k=\sum_r e^{-ikr}C(r)$. The exact self-energy follows from Eq.~(\ref{eq:6}): $\Sigma_k=\frac{T}{n_k}-\varepsilon_k+\mu$.

\setcounter{equation}{0}
\renewcommand{\theequation}{D\arabic{equation}}
\section*{A\lowercase{ppendix} D: S\lowercase{olving for equilibrium from the launch values}}
\label{Appendix_D}

\textit{Conserved launch quantities.}---
For an initial field configuration $\psi_m(0)$, the conserved power and Hamiltonian densities are

\begin{align}
a&=\frac{1}{M}\sum_m|\psi_m(0)|^2,\\
h&=\frac{1}{M}\left[
\sum_{m,n}h_{mn}\psi_m^*(0)\psi_n(0)
+\sum_m F(|\psi_m(0)|^2)
\right].
\label{eq:D1}
\end{align}

These quantities determine the equilibrium state within the local closure.

\textit{Dressed Rayleigh--Jeans constraints.}--
When the nonlinear self-energy is approximately independent of the mode index, the modal occupations take the dressed RJ form $n_k^{\rm dRJ}=T/(\varepsilon_{{k}}-\mu_{\mathrm{eff}})$ \cite{kabat2026optical}. The effective chemical potential is then fixed by the power constraint,

\begin{equation}
a=\frac{1}{M}\sum_k\frac{T}{\varepsilon_k-\mu_{\mathrm{eff}}},
\label{eq:D2}
\end{equation}
and the linear energy density is
\begin{equation}
h_{\mathrm{L}}^{\mathrm{dRJ}}(T,a)=\frac{1}{M}\sum_k\frac{\varepsilon_k T}{\varepsilon_k-\mu_{\mathrm{eff}}}.
\label{eq:D3}
\end{equation}
The nonlinear energy density follows from the local distribution, Eq.~(\ref{eq:10}); for the Kerr response,
\begin{equation}
h_{\mathrm{NL}}^{\mathrm{cl}}(T,a)=\frac{\chi a^2}{2}\,g_2\!\left(\frac{\chi a^2}{T}\right).
\label{eq:D4}
\end{equation}

\textit{Self-consistent temperature determination.}---
For a trial temperature $T$, Eq.~(\ref{eq:D2}) is first solved for $\mu_{\mathrm{eff}}<\varepsilon_{\mathrm{min}}$; Eqs.~(\ref{eq:D3}) and (\ref{eq:D4}) then give the linear and nonlinear energies, and the energy constraint, Eq.~(\ref{eq:14}), is solved for $T$. At fixed $a$, both terms on the right side of Eq.~(\ref{eq:14}) increase with $T$, so that a positive-temperature solution, when it exists, is unique. The modal occupations then follow from $n_k^{\mathrm{dRJ}}$, and the relation between the effective and thermodynamic chemical potentials follows from the approximately uniform self-energy:
\begin{equation}
\mu-\mu_{\mathrm{eff}}\simeq\bar{\Sigma}=\chi a\,g_2\!\left(\frac{\chi a^2}{T}\right).
\label{eq:D5}
\end{equation}
This prediction relies on the scalar dressed RJ approximation and is therefore applicable when the self-energy varies weakly across the occupied modes. The local closure supplies the nonlinear energy contribution, while the modal constraints account for the linear spectrum.

\setcounter{equation}{0}
\renewcommand{\theequation}{E\arabic{equation}}
\section*{A\lowercase{ppendix} E: B\lowercase{enchmark tables}}
\label{Appendix_E}

%We integrate Eq.~(\ref{eq:1}) with a second-order split-step scheme \cite{weideman1986split}, treating the linear step exactly in the supermode basis and the nonlinear step exactly in real space; both steps conserve the power exactly, and the energy is conserved to second order in the step size. The launched fields have uniform intensity, $\psi_m(0)=\sqrt{a}\,e^{i\phi_m}$, with random phases whose spatial correlation sets $h$. Observables are averaged over the propagation after an initial transient and over independent realizations (Table~\ref{tab_2}). 
Table~\ref{tab_1} tests the virial relation, Eq.~(\ref{eq:4}), on the exact solutions of Appendix~\hyperref[Appendix_C]{C}. Table~\ref{tab_2} lists the twelve simulated states (1D and 2D Kerr, 1D saturable), with their launch values $(a,h)$ and the linear energy at launch, in the simulation at equilibrium, and as predicted from the launch values by the local and Gaussian closures.

\begin{table}[h!]
\caption{Test of the optical virial relation on the exact 1D solution. The linear energy $h_{\rm L}$ obtained from the bond statistics of the transfer operator, Eq.~(\ref{eq:C3}), is compared with $T+\mu a-\langle AF'\rangle$ from the virial relation (virial); NL: nonlinearity. Kerr: $F'=\chi A$ with $\chi=1$. Saturable (Sat.): $F'=A/(1+A)$, i.e., $\chi=I_s=1$; $J=1$.}
\label{tab_1}
{\small
\setlength{\tabcolsep}{3pt}
\begin{ruledtabular}
\begin{tabular}{lrrrrrr}
 & & & & & \multicolumn{2}{c}{$h_{\rm L}$}\\
\cline{6-7}
NL & $\beta$ & $\mu$ & $a$ & $g_2$ & TIO & virial\\
\hline
Kerr & 1.0 & $-3.0$ & 0.3323 & 1.8465 & $-0.200897$ & $-0.200897$\\
Kerr & 0.2 & $-5.0$ & 0.8110 & 1.8274 & $-0.256572$ & $-0.256572$\\
Kerr & 0.05 & 2.0 & 4.6151 & 1.4681 & $-2.039811$ & $-2.039811$\\
Kerr & 3.0 & $-2.2$ & 0.2058 & 1.8228 & $-0.196679$ & $-0.196679$\\
Sat. & 1.0 & $-3.0$ & 0.3680 & 1.9287 & $-0.241689$ & $-0.241689$\\
Sat. & 0.2 & $-5.0$ & 0.9588 & 1.9610 & $-0.355126$ & $-0.355126$\\
Sat. & 3.0 & $-2.2$ & 0.2285 & 1.8831 & $-0.232442$ & $-0.232442$\\
\end{tabular}
\end{ruledtabular}
}
\end{table}

\begin{table}[h!]
\caption{Simulated states and predictions from the launch values $(h,a)$. Here, $a$ is the power density and $h$ is the conserved energy density of the launched field. The linear energy density $h_{\rm L}$ is reported at launch, in the simulation at equilibrium (sim.), and as predicted by the local and Gaussian (Gauss.) closures. A dash indicates that the Gaussian closure has no positive-temperature solution. System specifies the lattice and nonlinearity; Kerr: $\chi=1$; saturable (Sat.): $\chi=I_s=1$; $J=1$. Realizations: 100 (1D Kerr), 50 (2D Kerr), and 100 (1D saturable). 1D chains: 256 sites; 2D lattices: $32\times32$ sites.}

\label{tab_2}
{\small
\setlength{\tabcolsep}{2pt}
\begin{ruledtabular}
\begin{tabular}{cccrrrrr}
 & & & & \multicolumn{4}{c}{$h_{\rm L}$}\\
\cline{5-8}
State & System & $a$ & $h$ & launch & sim. & local & Gauss.\\
\hline
A  & 1D Kerr & 0.5 & 0.109    & $-0.016$ & $-0.128$ & $-0.129$ & $-0.141$\\
B  & 1D Kerr & 2.0 & $-1.336$ & $-3.336$ & $-3.565$ & $-3.564$ & --\\
C  & 1D Kerr & 5.0 & 12.404   & $-0.096$ & $-4.012$ & $-3.984$ & --\\
D  & 1D Kerr & 3.0 & $-1.313$ & $-5.813$ & $-5.885$ & $-5.876$ & --\\
E  & 1D Kerr & 1.0 & $-1.477$ & $-1.977$ & $-1.984$ & $-1.985$ & --\\
A2 & 2D Kerr & 0.5 & 0.130    & 0.005    & $-0.114$ & $-0.114$ & $-0.120$\\
B2 & 2D Kerr & 2.0 & 2.003    & 0.003    & $-1.374$ & $-1.375$ & $-1.997$\\
C2 & 2D Kerr & 4.0 & 8.015    & 0.015    & $-3.930$ & $-3.929$ & $-7.985$\\
D2 & 2D Kerr & 2.0 & $-5.676$ & $-7.676$ & $-7.762$ & $-7.784$ & --\\
S1 & 1D Sat. & 2.0 & 0.887    & $-0.014$ & $-0.189$ & $-0.189$ & $-0.190$\\
S2 & 1D Sat. & 6.0 & 3.875    & $-0.179$ & $-0.499$ & $-0.501$ & $-0.501$\\
S3 & 1D Sat. & 4.0 & $-5.258$ & $-7.649$ & $-7.694$ & $-7.750$ & $-7.918$\\
\end{tabular}
\end{ruledtabular}
}
\end{table}

%%%%%%%%%%%%%%%%%%%%%%%%%%%%%%%%%%%%%%%%%%%%%%%%%%%%%%%%%%%%%%%%%%%%%%%%%%%%%%%%%%%%%%%
\bibliography{bib_file.bib}
\end{document}